\documentclass[prl,twocolumn,superscriptaddress,showpacs]{revtex4}
\usepackage{epsfig,amsmath,amssymb,graphics,color,calc}

\newcommand{\be}{\begin{equation}}
\newcommand{\ee}{\end{equation}}
\newcommand{\ba}{\begin{eqnarray}}
\newcommand{\ea}{\end{eqnarray}}
\renewcommand{\phi}{\varphi}

\begin{document} 
\title{Anomalous structural evolution of soft particles: 
Equibrium liquid state theory}

\author{Hugo Jacquin}
\affiliation{Laboratoire Mati\`ere et Syst\`emes Complexes, UMR CNRS 7057,
Universit\'e Paris Diderot -- Paris 7, 10 rue Alice Domon et L\'eonie 
Duquet, 75205 Paris cedex 13, France}

\author{Ludovic Berthier}
\affiliation{Laboratoire des Collo{\"\i}des, Verres
 et Nanomat{\'e}riaux, Universit{\'e} Montpellier II and CNRS,
34095 Montpellier, France}

\date{\today}
\begin{abstract}
We use the hyper-netted chain approximation of liquid state theory
to analyze the evolution with density of the pair correlation function
in a model of soft spheres with harmonic repulsion.  
As observed in recent experiments on jammed soft particles, 
theory predicts an `anomalous' (nonmonotonic) 
evolution of the intensity of the first peak when density is increased at 
constant temperature. This structural anomaly is a direct consequence of 
particle softness, and can be explained from purely 
equilibrium considerations, emphasizing the generality
of the phenomenon. 
This anomaly is also predicted to have a non-trivial,
`${\cal S}$-shaped', evolution with temperature, as a
result of a competition between three distinct  effects, which we
describe in detail. Computer simulations support our predictions.
\end{abstract}

\pacs{05.20.Jj, 64.70.qd, 83.80.Fg}


\maketitle

\begin{figure}[b]
\includegraphics[width=8.5cm]{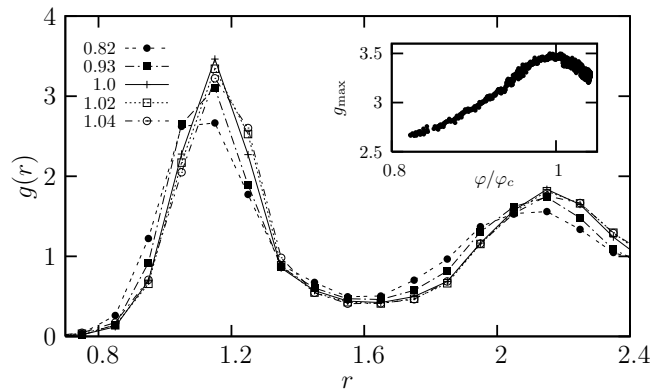}
\caption{\label{tap} Nonmonotonic evolution with increasing density 
of the height of the first peak in the pair correlation 
of a slowly swelling assembly of (soft) tapioca pearls. 
Volume fractions $\phi$ are normalized with respect to that 
of the maximum location at $\phi_c$. The inset shows the height 
of the first peak near $r \approx 1.1$ as a function
of $\phi/\phi_c$.
 These data are adapted from Ref.~\cite{tapioca}, 
and were kindly provided by X. Cheng.} 
\end{figure}

In very 
recent work~\cite{yodh,tapioca,tapioca2,xu}, the structure 
of soft repulsive particles was analyzed experimentally and numerically. 
Particular attention was paid to  
a nonmonotonic behaviour of the first peak of the 
pair correlation with increasing 
density, as shown in Fig.~\ref{tap}. 
In the limit where particles are infinitely hard, 
the height of the first peak 
diverges when approaching a maximal packing fraction 
where the pressure diverges~\cite{donev,leo}, and 
configurations with larger density 
cannot be explored. 
Thus, the nonmonotonic height of the pair correlation function
$g(r)$ was understood as a smooth, finite-temperature, signature 
of this jamming transition~\cite{yodh}.
While simulations study soft repulsive potentials
at finite temperatures~\cite{yodh,xu},
experiments introduce instead particle softness, which 
similarly allows for 
finite overlaps between particles. 
For instance, in the very elegant experiment of Ref.~\cite{tapioca}, 
the pair correlation function of 
a slowly swelling assembly of (compressible) tapioca pearls is 
followed, see Fig.~\ref{tap}.

These studies were performed in nonequilibrium conditions
following a preparation protocol 
which varies from one experiment to another.
Here, `nonequilibrium' means that particles rearrangements 
are rare during the course of an experiment, and do not
allow the system to sample equilibrium trajectories 
at any given state point, even at finite temperature.
The jamming transition cannot be probed  
at thermal equilibrium, because the system first hits 
a glass transition and cannot maintain ergodic 
behaviour at such large densities~\cite{ohern,zamponi,tom}. 
Thus, the structural evolution 
along particular nonequilibrium trajectories
does not result from ensemble averages
in the usual sense. When comparing two different state points, 
it is understood that a single configuration is transported
from one state point to another along a specific path. This experimental 
procedure suggests that sample preparation 
protocols might well play an important  role.

There are several questions left unanswered by these studies. 
First, the origin of the maximum is systematically attributed 
to an underlying $T=0$ jamming transition, and it is thus 
unclear how the phenomenon should be described 
theoretically. Second, the effects of sample preparation were
either not discussed, or claimed to be negligible~\cite{yodh}.
This seems to contradict recent reports that the location
of the jamming transition is sensitively protocol 
dependent~\cite{torquato1,torquato2,pinaki,hermes} 
a conclusion which should extend to finite temperatures as well. 
Third, the location of the maximum was predicted 
to obey a simple scaling behaviour near zero temperature~\cite{yodh}, 
as confirmed by simulations spanning a large temperature 
window~\cite{xu}. 
However, this prediction must break down when 
temperature is large enough to allow relaxation 
in the system, but this has not been explored.

In this paper, our aim is to gain theoretical insight 
into the behaviour of the pair correlation function of
dense assemblies of soft repulsive particles and to address
the three above questions. 
We suggest that the simplest starting point is equilibrium 
statistical mechanics, and we use the tools 
of liquid state theory to attack the problem. 
Although we know beforehand that our results
should eventually become inapplicable to the low temperature
regime where simulations and experiments fail
to probe equilibrium configurations, 
our simple approach provides useful insights, 
which we report in this short note.

We follow Refs.~\cite{yodh,xu} and study an assembly 
of $N$ soft repulsive particles interacting through 
harmonic repulsion:
\be
V(r_{ij} < 1) = \epsilon (1-r_{ij})^2,
\label{pair}
\ee
where $r_{ij}=|r_i - r_j|$ denotes the distance between
particles $i$ and $j$. 
Particles separated by $r_{ij}>1$ do not interact, $V(r_{ij}>1)=0$.
In the following, we report temperatures in units of $\epsilon$, 
and the particle diameter sets the unit lengthscale. 
We study the pair correlation function 
defined by
\be
g(r) = \frac{1}{\rho N} 
\sum_{i \neq j} \left\langle \delta (r - r_{ij}) \right\rangle,
\ee
where the brackets indicate an ensemble average at thermal
equilibrium, and $\rho = N/V$ is the number density.
We use the hyper-netted chain (HNC) closure relation 
for the pair correlation function~\cite{hansen}, 
 \be
g(r) = \exp[ -\beta V(r) + g(r) - 1 -c(r) ],
\label{hnc}
\ee
for the pair potential $V(r)$ in Eq.~(\ref{pair}). 
We have defined $\beta = 1/T$, where $T$ is the temperature.
In Eq.~(\ref{hnc}),  $c(r)$ is the direct correlation function defined 
through the Ornstein-Zernike equation:
\be
g(r)-1 = c(r) + \rho \int dr' c(r-r') (g(r')-1).
\label{OZ}
\ee
Although HNC is known to become quantitatively inaccurate in the hard sphere 
limit at the large densities investigated here~\cite{hansen}, 
it is a convenient
tool to study the interplay between temperature and dynamics 
in soft repulsive spheres. For the relatively large temperatures
studied below, we shall find that HNC in fact 
compares quite well with simulations, at least qualitatively.

We solve the HNC closure relation in Eq.~(\ref{hnc}) numerically. 
To this end, we discretize $g(r)$ between $r=0$ and $r=32$, 
using $2^{15}$ points and use a standard Picard iterative
procedure, as described in more details in Ref.~\cite{hugo}.
As a result, we obtain $g(r)$ over a broad range of temperatures,
$T \in [0.00076, 0.5]$ and volume fractions, $\phi = \pi \rho / 6 
\in [0.5, 1.2]$. 

\begin{figure}
\includegraphics[width=8.5cm]{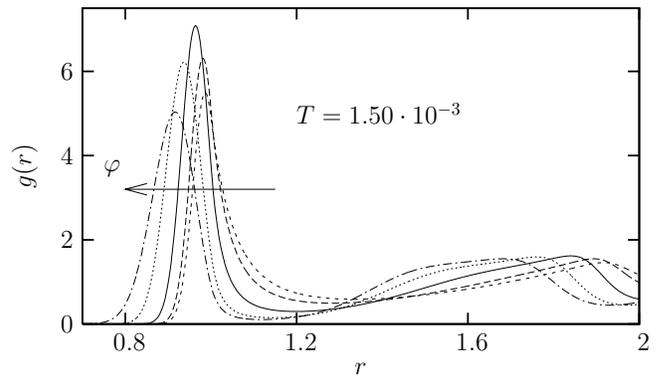}
\caption{\label{vestige} Nonmonotonic evolution of the 
height of the first peak in the pair correlation $g(r)$, 
obtained using the hyper-netted chain approximation. 
The temperature $T$ is fixed, and the volume fraction 
increases from right to left with 
$\phi=0.600$, 0.663, 0.777, 0.900, and 1.00.}
\end{figure}

Our first result is shown in Fig.~\ref{vestige}, 
where the pair correlation of harmonic spheres 
at constant temperature, $T=1.5 \cdot 10^{-3}$, is
shown for volume fractions increasing from $\phi=0.60$
up to $\phi=1.0$. While the position of the first peak 
shifts to smaller distances as a result 
of the compression, the height of the first peak 
displays a nonmonotonic evolution. For this temperature, 
it increases up to $\phi \approx 0.777$, and then decreases 
for larger volume fractions. This evolution is very similar
to the one reported in Fig.~\ref{tap}. 
In Fig.~\ref{vestige}, it can be seen that the second 
peak of $g(r)$ also displays a nonmonotonic behaviour, while 
we find the subsequent peaks do not. Remarkably, the 
same observations were reported in experiments~\cite{tapioca}.

In the context of jamming and glassy physics, this result is 
unexpected, since in simple glass-forming liquids, modelled for 
instance as Lennard-Jones fluids, the first peak of 
$g(r)$ typically increases with density at large density, contrary to 
the system of harmonic spheres, where it instead decreases, see
Fig.~\ref{vestige}.

This result shows that a nonmonotonic 
structural evolution of $g(r)$ exists even at thermal
equilibrium and not only in the 
nonequilibrium iso-configurational conditions investigated 
in Refs.~\cite{yodh,tapioca,tapioca2,xu}. 
Therefore, we conclude that equilibrium concepts can be used
to understand its origin and we suggest that the mechanism 
behind the behaviour in Fig.~\ref{vestige} is quite generic
and not specifically due to an underlying nonequilibrium jamming 
transition at $T=0$.
Indeed, we find in the literature several observations 
of similar anomalies observed in equilibrium 
conditions in systems of soft repulsive particles,
for instance for Hertzian spheres~\cite{frenkel} or in 
the Gaussian core model~\cite{likos,truskett}.
To our knowledge, the close 
correspondence between nonequilibrium observations 
near the jamming transition, and these earlier reports 
at thermal equilibrium was not  noted before. 

Physically, the anomaly is a direct consequence of 
particle softness~\cite{frenkel,truskett,likos,louis}. The system 
behaves as an effective hard sphere fluid at sufficiently low
volume fraction where it can easily minimize energy by
avoiding particle overlaps. In this regime, the height
of the peak increases with $\phi$, just as it does in 
a hard sphere fluid. When the volume fraction increases, however,
it becomes harder, and thus entropically less favourable, 
to avoid particle overlaps as the system must find 
more efficient ways to pack the particles. Thus, the system 
may gain free energy by increasing the
disorder at the expense of allowing particle overlaps, which 
are energetically not very costly for soft repulsions 
such as the one in Eq.~(\ref{pair}).  
In  this regime, disorder increases when increasing $\phi$, 
and this broadens the first peak of $g(r)$ and decreases its height.
We note that similar competitions are invoked 
to explain an even larger set of anomalous thermodynamic behaviours, 
as observed for instance in widely studied network-forming 
liquids such as silica~\cite{silica} and water~\cite{water,water2}.

\begin{figure}
\includegraphics[width=8.5cm]{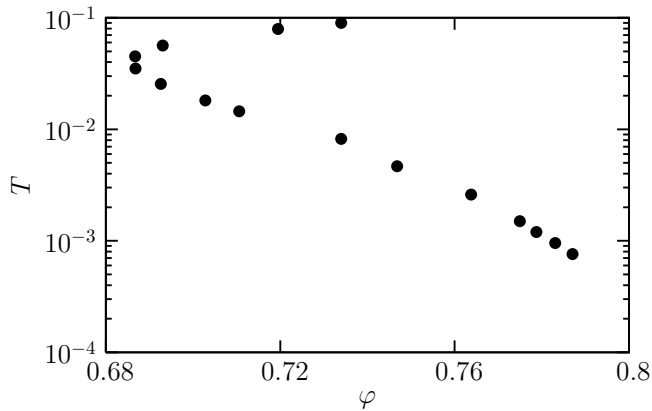}
\caption{\label{locus} Location of the 
maximum of the first peak of $g(r)$ in a temperature ($T$) --
volume fraction ($\phi$) phase diagram. This location 
is itself a nonmonotonic function of temperature.
In particular in the low $T$ regime, 
it shifts to large $\phi$ when $T$ decreases, as the system 
is able to find more efficient ways to pack 
non-overlapping particles.}
\end{figure}

Since we attribute the anomaly to a competition 
between energy and entropy, the outcome 
of this competition can be expected to have an interesting 
evolution with temperature. Thus, 
we now turn to the influence of temperature on the 
nonmonotonic behaviour described in Fig.~\ref{vestige}. 
In Fig.~\ref{locus}, we follow
the locus of the maximum of $g(r)$ in a 
$(T,\phi)$ phase diagram, as predicted from the HNC approximation.
This figure confirms that equilibrium theory 
predicts a fairly nontrivial temperature evolution of the 
anomaly location, which is itself a 
nonmonotonic function of the temperature, with a crossover 
near $T \approx 0.04$. For $T > 0.04$ the anomaly 
shifts to larger density for increasing temperature, while the opposite 
behaviour is found for $T<0.04$. 
 
The low temperature behaviour, $T < 0.04$, is readily understood
on the basis of the equilibrium argument laid out above. 
When $T$ decreases, the entropic penalty to find efficient 
packing of the particles contributes less to the free energy, and it is 
preferable to avoid the energetically costly 
particle overlaps up to larger volume fractions. 
Thus, from equilibrium considerations one would naturally
predict qualitatively the low temperature behaviour 
reported in Fig.~\ref{locus}. It reflects 
the physical intuition that thermal annealing should 
help the system to  `solve the packing problem'
more efficiently.
 
Temperature has a second, competing effect which 
explains the high temperature behaviour in Fig.~\ref{locus}.
When $T$ gets larger, particles acquire more kinetic
energy, and they can more easily overlap. Thus the `effective'
particle diameter gets reduced by a simple thermal effect when 
$T$ increases~\cite{yodh,tom}, 
and this effect must be compensated by an increase
of the volume fraction. 
Therefore, temperature has two opposite effects which combine to yield the 
behaviour reported in Fig.~\ref{locus}, and explain the 
crossover near $T \approx 0.04$.

What happens if temperature is decreased further?
We mentioned in the introduction 
that our equilibrium calculations eventually become inapplicable 
since a glass transition should take place when the 
structural relaxation time gets very large. In a simulation
or an experiment, there will be a temperature below 
which the system will eventually be frozen into some
amorphous configuration~\cite{tom}. It is possible to describe 
this glass physics using liquid state theory, but 
this requires considerably more 
advanced techniques~\cite{zamponi,replica1,replica2,hugo2}.

Still, a number of conclusions naturally follow from the above 
equilibrium results.
First, at low but finite temperatures, 
the system is deep into the glass region. Although it cannot 
relax to equilibrium on accessible timescales, it attempts to do so,
albeit at a very slow pace. Thus, the physical explanation
given above for the very existence of the structural anomaly 
should still apply in this regime, which is the one 
explored  in Refs.~\cite{yodh,xu}. 
The structural anomaly should thus be `frozen in' by the 
glass transition, which involves no important change
of the structure.
Second, we have shown above that the location 
of the maximum is ruled by two competing effects. In the glass phase, 
only the second one remains effective, as the system is not able
to explore different configurations efficiently and find better
particle packings. Thus, the evolution of the anomaly at low
temperature should change again as the glass transition is crossed, 
and revert to the high temperature behaviour of Fig.~\ref{locus}. 
Overall, thus, we predict that the locus of the structural anomaly
displays three regimes due to the competing effects of 
(i) thermal fluctuations reducing the effective
particle diameter, (ii) the entropy / energy competition 
favouring more efficient packing at low temperatures, 
(iii) the intervening glass transition. 
In a $(T, \phi)$ phase diagram, the data should thus 
follow a nontrivial `$\cal S$-shape'. This theoretical prediction
is our main new result.

\begin{figure}
\includegraphics[width=8.5cm]{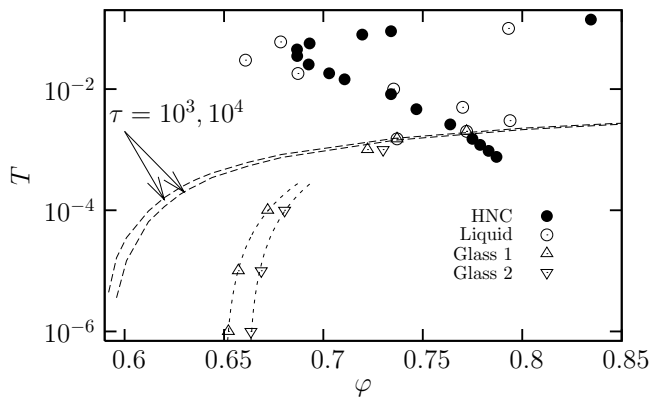}
\caption{\label{locus2} Location of the 
maximum of the first peak of $g(r)$ in a temperature 
volume fraction phase diagram from HNC (filled circles),
equilibrium simulations (open circles), nonequilibrium 
glass configurations (open triangles). The glass transition 
are defined as iso-relaxation timescales lines, with two 
different large values, from Ref.~\cite{tom}. The location of the maximum
follows a nontrivial ${\cal S}$-shape, as predicted
theoretically.}
\end{figure}

We confront these theoretical considerations to numerical 
simulations in Fig.~\ref{locus2}. We report the HNC prediction 
for the location of the structural anomaly from Fig.~\ref{locus}
and compare it
to the outcome of molecular dynamics simulations at thermal equilibrium.  
To obtain these data points, we made use of the extensive
set of data of Ref.~\cite{tom}, where a 50:50 binary mixture 
of harmonic spheres with size ratio 1.4 was studied using molecular dynamics.
A mixture is needed in the simulation to avoid 
crystallization. 
When necessary, we performed additional
simulations to explore a broader range of state points,
using the same techniques. We use a large system size, $N=8000$, 
where finite size effects are negliglible both at equilibrium~\cite{tom}
and at very low temperatures~\cite{ohern,pinaki}. 

We find a very good qualitative agreement
between the equilibrium behaviour predicted 
using HNC and the outcome of computer simulations at sufficiently 
high temperatures, $T > 0.003$, confirming
that HNC performs quite well for harmonic spheres
in this range of densities 
and temperatures. The nonmonotonic temperature 
behaviour and the location of the crossover near $T\approx 0.04$ 
are well reproduced. Quantitative agreement 
cannot be expected, as size polydispersity was not taken into
account in our theoretical calculations. 

We then report glass transition lines, defined from the temperature
where the structural relaxation time, $\tau$, becomes larger than 
some prescribed value. We report two lines corresponding to two 
large values, $\tau = 10^3$ and $\tau = 10^4$, in the reduced time
units used in molecular dynamics simulations~\cite{tom}. 
It is very difficult to maintain thermal equilibrium below these 
lines in computer simulations.
These glass lines meet the locus of structural anomaly 
near $\phi \approx 0.8$ and  $T \approx 0.0025$, 
which shows that the maximum of $g(r)$ cannot
be observed in equilibrium conditions in computer simulations
below $T=0.0025$. This also suggests that both 
phenomena are not directly connected. 

At even lower temperatures, it becomes ambiguous to 
locate a unique  maximum of $g(r)$, as this should 
in principle depend on the way the glass 
configurations have been prepared. To illustrate this difficulty, and 
therefore the 
nonequilibrium nature of the low temperature behaviour,
we numerically 
prepared two glasses of harmonic spheres: 
`Glass 1' is very rapidly quenched from a high 
temperature to well below the glass transition, while 
`Glass 2' is very slowly annealed across $T_g$. 
For both glasses, we track the location of the $g(r)$ maxima
down to very low temperatures, see Fig.~\ref{locus2}. 
As discussed above, these data confirm the 
behaviour predicted at low temperature when no structural 
relaxation can occur. The location of the maximum
shifts to lower volume fraction when temperature decreases.
This is precisely the regime of temperature described in previous 
work~\cite{yodh,xu}, and our results for `Glass 1' 
agree with those earlier reports.
In Fig.~\ref{locus2}, we report the low temperature 
scaling behaviour deduced from considering the effect of thermal
fluctuations alone, $T \sim (\phi - \phi_c)^2$, where the exponent 
`2' is the same as in the pair potential (\ref{pair})
and $\phi_c$ denotes the location of the jamming transitions in the 
$T=0$ limit. These lines separate in the glass phase
a region where the glass is dominated by repulsion at low volume fraction, 
from a region where particles are compressed at large volume fraction, 
and the system resembles a compressed emulsion. 

The nonequilibrium nature of the jamming transition
is illustrated from the difference between 
our two glasses in Fig.~\ref{locus2}. They both display a line of structural 
anomaly, but these two lines are distinct and 
define jamming transitions at distinct densities, 
$\phi_c=0.65$ and $\phi_c=0.662$. 
The dependence of $\phi_c$ on the preparation protocol
in the $T=0$ limit has been discussed 
before~\cite{torquato1,torquato2,pinaki,hermes}. The data 
in Fig.~\ref{locus2} show that these considerations apply 
to finite temperatures as well, and to the structural 
anomaly in particular.
Therefore, this figure illustrates the fact that nonequilibrium jamming 
transition lines are as ambiguously defined 
as glass transition lines, since they both strongly depend 
upon operational definitions~\cite{pinaki,kurchan,torquato3}. 
These two concepts can only become protocol-independent in theoretical 
calculations~\cite{zamponi,hugo2}.

The crossover between the very low temperatures investigated in
previous work, and the equilibrium regime studied here is also
interesting. In the regime $T \approx 0.001 - 0.003$, the system
is an aging glass at large density, but a fluid at low density. 
For the fluid state points the first peak of $g(r)$ increases 
with $\phi$ and the extrapolated maximum is at very large $\phi$. 
For the aging glasses, however, the peak decreases with 
$\phi$ as the glass maximum should follow the low temperature trend.
Thus, these two tendencies produce a maximum when the glass transition 
is crossed, and  the glass and jamming lines in Fig.~\ref{locus2} `kiss' 
in this temperature regime. 
Of course, the exact location of the maximum
depends on the aging time.  This effect 
was not reported before~\cite{yodh,xu}, as the structure does 
not age very much at the very low temperatures studied in these earlier
reports, $T /T_g \approx 1/1000 -1/10$.  
 
It is interesting to remark the close similarity 
between the effect of glass annealing in nonequilibrium 
situations, and the effect of decreasing the temperature at equilibrium:
they both shift the maximum of $g(r)$ to larger $\phi$ for the same 
physical reason. This correspondence emphasizes once more 
that the physics of the structural anomaly described in this paper 
is in fact deeply rooted in equilibrium concepts. 
The $T \to 0$ limit is no exception, but it makes 
the distinction  between repulsive and compressed regimes
sharper, as the energy of the ground state of the system 
is either 0 or positive~\cite{ohern}.  

To conclude, we have used a simple equilibrium theoretical approach to
study the existence and location of a structural anomaly 
reported in recent experiments in soft repulsive spheres
in nonequilibrium conditions.
We have suggested that the structural anomaly reported 
in these articles can be explained using purely equilibrium 
concepts, and several aspects of its behaviour 
are indeed well described using liquid state theory. 
The very existence of an anomaly relies on a standard competition 
between entropy and energy in the free energy of these 
very soft particles, 
and we have suggested that the anomaloy belongs to a much 
broader class of similar anomalies studied in the literature. 
In the present case, it directly results from 
the extreme softness of particles described by pair
potentials such as Eq.~(\ref{pair}). 

For standard
models describing atomic liquids, such as Lennard-Jones 
potentials, the repulsive core is much steeper, and they are thus governed
by a different physics upon compression. Indeed, the large density
limit of both types of systems is very different, as the harmonic 
spheres behave as an ideal gas in this limit~\cite{likos}.
In the context of glassy physics, a recent quantitative 
comparison between harmonic spheres and Lennard-Jones models 
of the glass transition has also confirmed the different nature 
of their glassy dynamics and of its evolution with 
density~\cite{gilles}.

Additionally, we have discussed
three competing effects of the temperature which control
the location of the anomaly in the phase diagram, 
and predicted a nontrivial `${\cal S}$-shape', confirmed 
by numerical simulations. We have finally shown 
that the anomalous structural behaviour of
soft particles discussed in Refs.~\cite{yodh,tapioca,tapioca2,xu} 
in fact occurs over a broad, continuous range 
of densities and temperatures, as it is directly 
affected by sample preparation effects.

In future work~\cite{hugo2}, we plan to extend the present liquid 
state calculations to describe analytically the properties
of dense systems of soft particles down to very low 
temperatures, where ergodicity breaking must 
explicitely be taken into account.
  
\acknowledgments
We thank G. Biroli, L. Cipelletti, W. Kob, G. Szamel, and F. Zamponi
for useful feedback and constructive criticisms about this work.
We thank X. Cheng for providing his experimental 
data from Ref.~\cite{tapioca}.
H. Jacquin acknowledges financial support from Capital Fund Management (CFM)
Foundation, and from the LCVN in the early stages of this work.
L. Berthier is partially supported by the ANR Dynhet.


\begin{thebibliography}{99}

\bibitem{yodh}
Z. Zhang, N. Xu, D. T. N. Chen, P. Yunker, 
A. M. Alsayed, K. B. Aptowicz, P. Habdas, A. J. Liu, 
S. R. Nagel, and A. G. Yodh, Nature {\bf 459}, 230 (2009). 

\bibitem{tapioca2} X. Cheng, 
arXiv:0905.2788 (2009).

\bibitem{tapioca} X. Cheng,
arXiv:0911.1943 (2009).

\bibitem{xu} N. Xu, 
arXiv:0911.1576 (2009).

\bibitem{donev}
A. Donev, S. Torquato, and F. H. Stillinger,
Phys. Rev. E {\bf 71}, 011105 (2005).

\bibitem{leo}
L. E. Silbert, A.  J. Liu,
and S. R. Nagel, Phys. Rev. E {\bf 73}, 041304 (2006).

\bibitem{ohern}
C. S. O'Hern, S. A. Langer, A. J. Liu, and S. R. Nagel, Phys. Rev.
E {\bf 68}, 011306 (2003).

\bibitem{zamponi} G. Parisi and F. Zamponi,
Rev. Mod. Phys. (in press); arXiv:0802.2180.

\bibitem{tom}
L. Berthier and T. A. Witten,
EPL {\bf 86}, 10001 (2009); 
Phys. Rev. E {\bf 80} 021502 (2009).

\bibitem{torquato1} A. Donev, S. Torquato, F. H. Stillinger, and R. Connelly,
Phys. Rev. E {\bf 70}, 043301 (2004).


\bibitem{torquato2} A. Donev, F. H. Stillinger, and S. Torquato,
J. Chem. Phys. {\bf 127}, 124509 (2007).

\bibitem{pinaki} 
P. Chaudhuri, L. Berthier, and S. Sastry,
arXiv:0910.0364 (2009).

\bibitem{hermes} M. Hermes and M. Dijkstra,
arXiv:0903.4075.

\bibitem{hansen}
J. P. Hansen and I. R. McDonald, {\it Theory of Simple Liquids}
(Elsevier, Amsterdam, 1986).

\bibitem{hugo}
L. Berthier, E. Flenner, H. Jacquin, and G. Szamel,
arXiv:0912.1738.

\bibitem{frenkel} J. C. Pamies, A. Cacciuto, and D. Frenkel,
J. Chem. Phys. {\bf 131}, 044514 (2009).

\bibitem{likos}
A. Lang, C. N. Likos, M. Watzlawek, and H. L\"owen, 
J. Phys.: Condens. Matter {\bf 12}, 5087 (2000). 

\bibitem{truskett}
W. P. Krekelberg, T. Kumar, J. Mittal, J. R. Errington, and
T. M. Truskett, Phys. Rev. E {\bf 79}, 031203 (2009).

\bibitem{louis} A. A. Louis, P. G. Bolhuis, and 
J. P. Hansen, Phys. Rev. E {\bf 62}, 7961 (2000).

\bibitem{silica}
M. S. Shell, P. G. Debenedetti, and A. Z. Panagiotopoulos,
Phys. Rev. E {\bf 66}, 011202 (2002).

\bibitem{water}
J. R. Errington and P. G. Debenedetti, Nature {\bf 409}, 318 (2001).

\bibitem{water2}
W. P. Krekelberg, J. Mittal, V. Ganesan, and T. M. Truskett,
Phys. Rev. E {\bf 77}, 041201 (2008). 


\bibitem{replica1} R. Monasson, Phys. Rev. Lett. {\bf 75},
2847 (1995).

\bibitem{replica2} M. M\'ezard and G. Parisi,
Phys. Rev. Lett. {\bf 82}, 747  (1999).

\bibitem{hugo2} H. Jacquin et al. (in preparation).

\bibitem{kurchan} F. Krzakala and J. Kurchan, 
Phys. Rev. E {\bf 76}, 021122 (2007).

\bibitem{torquato3} S. Torquato, T. M. Truskett, and P. G.
Debenedetti, Phys. Rev. Lett. {\bf 84}, 2064 (2000).


\bibitem{gilles} L. Berthier and G. Tarjus, 
Phys. Rev. Lett. {\bf 103}, 170601 (2009).

\end{thebibliography}
\end{document}